\documentclass[%
 reprint,
 amsmath,amssymb,
 aps,
]{revtex4-1}

\usepackage{graphicx}
\usepackage{dcolumn}
\usepackage{bm}
\usepackage{amsmath}
\usepackage{hyperref}
\usepackage[mathlines]{lineno}
\usepackage{esint}
\usepackage{csquotes}
\usepackage{xcolor}

\begin{document}


\title{Quantum Electromagnetic Rate Theory of the Electron and the Meaning of the Fine-Structure Constant}
\thanks{The physical analysis developed here, although self-consistent, requires the acceptance of a reinterpretation of the meaning of the Planck constant as an electromagnetic phenomenon (see Eq.~\ref{eq:h}), which was demonstrated in a separate work~\citep{Bueno-2023} using Maxwell's equations for the description of the closed orbital dynamics of the electron in free space.}%

\author{Paulo Roberto Bueno}
 \email{paulo-roberto.bueno@unesp.br}
\affiliation{Institute of Chemistry, Department of Engineering, Physics and Mathematics, S\~ao Paulo State University, Araraquara, S\~ao Paulo State, Brazil\\
}

\date{\today}

\begin{abstract}
In a previous work, the meaning of the Planck constant $h = \left( e^2 / 2 \alpha \right) \sqrt{\mu_0 / \epsilon_0}$, accomplished by solving Maxwell's electrodynamics laws with specific electric $1 / \tau_C = 1 / R_q C_q$ and magnetic $1/ \tau_L = R_q / L_q$ quantum rates for the ground-state dynamics, was reinterpreted. $R_q$, $C_q$ and $L_q$ are the resistance, capacitance and inductance quantum of the ground-state dynamics, respectively. Based on this quantum electromagnetic rate approach, here it is demonstrated that the intrinsic massless character that complies with Dirac quantum electrodynamics of the electron in its ground-state energy level is a consequence of a quantum electromagnetic phase coherence between $\tau_C$ and $\tau_L$ time constants of the oscillatory motion. The quantum mechanical uncertainties associated with $h$ are interpreted to be a consequence of perturbing the inherent electromagnetic phase coherence of the ground state with the loss of half of a byte of electromagnetic information per ``experimental'' perturbation, with the fine-structure constant $\alpha = \sqrt{\pi / 2 \left( \tau_L / \tau_C \right)} \sim 1/137$ playing a prominent role in the phenomenon. 

\end{abstract}

\maketitle


\section{\label{sec:introduction}Introduction}

In a previous work~\citep{Bueno-2023}, it was demonstrated that the ground-state energy $E_{gs}$ of an electron is associated with a particular electromagnetic character that complies with the relativistic massless Fermionic dynamics of the electron~\citep{Novoselov-2005}, that is, $ E_{gs} = \textbf{p} \cdot \textbf{c}_*$, where \textbf{c}$_*$ is the vector representation of the Fermi velocity $c_*$ and $\textbf{p} = \hbar \textbf{k}$ is the momentum vector, as predicted by the Dirac equation~\citep{Dirac-1928}. Additionally, it was noted that this particular electromagnetic description of the electron within closed orbital dynamics, remarkably, conforms with Maxwell's equations.

The analysis was a consequence of assuming a specific quantum rate $\nu_e = e^2 / g_e h C_q$ definition of the closed motion of the elementary charge of the electron, where $e$ is the elementary charge, $g_e$ is the electric field degeneracy of electron and equals to 2, $h$ is the Planck constant, and $C_q$ the capacitance quantum of the ground-state energy level. The energy $E_{gs} = h \nu_e = e^2 / g_e C_q$ (where $h \nu_e \sim$ 13.6 eV) associated with the rate $\nu_e$ was consistent with that predicted by Bohr~\citep{Kragh-2012}, whenever the Bohr radius $r_b = 5.29 \times 10^{-2}$ nm is taken into account in the definition of $C_q$ (see Eq.~\ref{eq:C_q-GS}).  

Furthermore, by following the quantum rate theoretical premise that establishes a correlation between $C_q$ and the conductance quantum $G_0 = 1/R_q = g_s e^2/h$ ($g_s$ is the spin degeneracy of the electron that equals to 2), not only was the above-described rate $\nu_e = e^2/g_e h C_q$ defined but, additionally, a magnetic quantum rate frequency of $\nu_m = R_q / L_q = g_s e^2 L_q/h$ was also defined. In the latter definition of $\nu_m$, note that $L_q$ refers to the quantum inductance. (Details about the definition and meaning of both $C_q$ and $L_q$ will be given later during the analysis of Eqs. \ref{eq:C_q-GS} and \ref{eq:L_q-GS}.) 

The electric component of the motion associated with $\nu_e$ was demonstrated~\citep{Bueno-2023} to be induced by the inherently closed movement of the elementary charge around the nuclei, in such a way that $E_{gs} = \textbf{p} \cdot \textbf{c}_* = e\oint_{\scriptstyle\partial\,\Sigma} \textbf{E} \,d\textbf{l} = \oint_{\scriptstyle\partial\,\Sigma} \textbf{F}_{i} \,d\textbf{l}$, where $\epsilon_i = \oint_{\scriptstyle\partial\,\Sigma} \textbf{E} = e/g_eC_q$ is an electromotive force induced by the time variation of the magnetic flux quantum $\Phi_B = h/e$ in synchrony to the period of time $\tau_C = R_qC_q$ and in agreement with Faraday's law. Therefore, $\oint_{\scriptstyle\partial\,\Sigma} \textbf{F}_{i} \,d\textbf{l}$ is the closed work resulting from an induced energy of $e\varepsilon_i$. 

Using above-defined electric $\nu_e$ and magnetic $\nu_m$ quantum frequencies it was not only possible to solve Maxwell's electrodynamics equations in compliance with Dirac quantum relativistic electrodynamics~\citep{Dirac-1928}, but additionally, this quantum electromagnetic rate approach conducted to a reinterpretation of the meaning of the Planck constant, in a manner that reveals its electromagnetic character, such as $ h = g_s e^2 / 4\alpha \sqrt{\mu_0/\varepsilon_0}$ or (by noting that $g_s$ equals to 2)

\begin{equation}
 \label{eq:h}
	h = \frac{e^2}{2\alpha}\sqrt{\frac{\mu_0}{\varepsilon_0}},
\end{equation}

\noindent where $\mu_0$ and $\epsilon_0$ are, respectively, the magnetic and electric permittivities of free space, $\alpha \sim 1/137$ is the fine-structure (or Sommerfeld) constant~\citep{Sommerfeld-1916}.

In Eq.~\ref{eq:h}, the fundamentally electromagnetic nature of $h$ is envisioned as being expressed solely in terms of fundamental constants of classic electromagnetism (i.e., $e$, $\mu_0$, and $\epsilon_0$, with the exception of $\alpha$), which is a dimensionless physical constant that quantifies the strength of the electromagnetic interaction between elementary charged particles~\citep{Gerlitz-2022, MacGregor-2007}. Note also that $\alpha$ is a fundamental constant of Standard Model of physical particles and the precision with it can be measured has been tremendous~\citep{Holger-Muller-2018}. 

It must be emphasized that, although $\alpha$ has an electromagnetic physical character, it is not \textit{per se} a component of classical electrodynamics~\citep{ED-Griffiths-1987}, but it is an important component of relativistic quantum dynamics~\cite{Feynman-1985}. Therefore, the meaning of $h$ brought by Eq.~\ref{eq:h} suggests pondering whether the origin of quantum leaps of energy observed at the atomic scale is fundamentally related to $\alpha$ instead of $h$. Actually, $\alpha$ has been an equivalent mystery~\cite{Gerlitz-2022, MacGregor-2007} in phenomena such as that observed for the quantum leaps of energy commonly attributed to $h$ in nonrelativistic quantum mechanics~\citep{Dirac-1928, Feynman-1985}. $\alpha$ is so important for quantum electrodynamics that Feynman stated that~\citep{Feynman-1985} 

\begin{displayquote}
``We know what kind of a dance to do experimentally to measure this number ($\sim$ 0.00729 $\dots$) very accurately, but we do not know what kind of dance to do on the computer to make this number come out -- without putting it in secretly!"
\end{displayquote} 

As stated thus far, it is important to recall that the classical electromagnetic theory~\citep{ED-Griffiths-1987} cannot address the stability of electron motion in the ground state because the orbiting electrons are constantly changing directions in a way that electromagnetic radiative emission is expected with a consequent continuous loss of energy with a bare predicted lifetime at the best of 10$^{-12}$ s. 

However, remarkably, quantum electromagnetic rate theory~\citep{Bueno-2023} not only complies with Maxwell's equations~\citep{ED-Griffiths-1987} of classical electrodynamics but also with Dirac relativistic electrodynamics~\citep{Dirac-1928}, because of the meaning of $\alpha$ provided by a quantum electromagnetic rate interpretation of the phenomenon, as will be demonstrated later. Therefore, quantum rate electrodynamics resembles phenomena involving magnetically trapped charges, but with a coherent dynamics between the electric and magnetic properties of the electron~\citep{Reens-2017}.  

Presently, to advance the interpretation of the electrodynamics of the ground state, the goal is to demonstrate the meaning of $\alpha$ within quantum electromagnetic rate theory~\citep{Bueno-2023} and discuss the consequences of these findings to our current comprehension of quantum electrodynamics. Therefore, it will be shown that $\alpha$ is related to $1 /\tau_C$ and $1 / \tau_L$ quantum rates, with $\tau_C = R_q C_q$ and $\tau_L = L_q / R_q$ referring to the characteristic time periods associated with the electric and magnetic dynamics of the electron in its relativistic ground-state orbital dynamics. The results are interpreted in view of an inherently quantum electromagnetic coherence existing between the electric and magnetic rate dynamics with $\alpha$ governing the phenomenon.

\section{\label{sec:EDP-Coherence} Quantum Electromagnetic Phase Coherence}

Akin to the resistance quantum~\citep{Imry-Landauer-1999} $R_q = h/g_s e^2$ defined as the reciprocal of the conductance quantum $G_0 = g_s e^2/h$, equivalently, it is possible to define capacitance $C_q$ and inductance $L_q$ phenomena for the ground state. This enables us to perform a quantum circuit analysis of the electromagnetic coherence existing in the ground state (see Section~\ref{sec:circuit} for more details). These definitions of $C_q$ and $L_q$ can be introduced by noting that $\tau_C = R_q C_q$ and $\tau_L = L_q / R_q$. Substituting $h$, as it is accounted for in Eq.~\ref{eq:h}, in the definition of $R_q = h/g_s e^2$ gives

\begin{equation}
 \label{eq:R_q-magnetic}
	R_q = \frac{1}{4\alpha} \sqrt{\frac{\mu_0}{\epsilon_0}} = \frac{Z_0}{4\alpha},
\end{equation}

\noindent demonstrating the intrinsic electromagnetic character of $R_q$, where $Z_0 = \sqrt{\mu_0/\epsilon_0}$ is known as the characteristic impedance of free space. Therefore, from Eq.~\ref{eq:R_q-magnetic}, it can be anticipated the adiabatic nature of $R_q$ that will be considered in Section~\ref{sec:circuit}. It must be stressed that $R_q$ is not a function of the radius $r$ of the orbital motion, as is the case of both $C_q$ and $L_q$ circuit elements, respectively, defined as

\begin{equation}
 \label{eq:C_q-GS}
	C_q = 2\lambda \epsilon_0
\end{equation}

and

\begin{equation}
 \label{eq:L_q-GS}
	L_q = \frac{\mu_0 \lambda}{4\pi},
\end{equation}

\noindent where $\lambda = 2 \pi r$ is the wavelength of the orbit of radius $r$ associated with the magnetically induced energy $e\varepsilon_i$~\citep{Bueno-2023}. This definition of $L_q = \mu_0 \lambda / 4 \pi = \mu_0 r / 2$ correlates with the meaning of Maxwell's induction law (see Eq.~\ref{eq:Maxwell-law}) defined in a previous work~\citep{Bueno-2023}, which, if rewritten in its simple form, leads to $B \lambda = \mu_0 i_L$ (see the meaning of $i_L$ in Section \ref{sec:Maxwelling-equations}). By noting now that $\Phi_B = B (\pi r^2) = B \lambda r / 2$, where $\pi r^2$ is the area through which the magnetic flux operates, the term $B \lambda$ is identified as $2 \phi_B / r$ and thus, by equating $2 \Phi_B / r = \mu_0 i_L$, the definition of $L_q$ is promptly obtained by comparing the latter result rearranged as $\Phi_B = \left( \mu_0 r /2 \right) i_L$~\footnote{Note also the definition of inductance $L$ as the ratio of $\Phi_B$ per unit of electric current $i$ such as $L = \Phi_B/i$~\citep{ED-Griffiths-1987}.} to a simpler form of Maxwell's equation $\Phi_B = \mu_0 i_L$ resulting from the quantum electromagnetic setting of Maxwell's laws~\citep{Bueno-2023}.

By using the definitions brought by Eqs. \ref{eq:C_q-GS} and \ref{eq:L_q-GS} and of $R_q$ in Eq.~\ref{eq:R_q-magnetic}, the characteristic time constants $\tau_C$ and $\tau_L$ of the closed orbital motion can be, respectively, written as

\begin{equation}
 \label{eq:tauC}
	\tau_C = R_q C_q = \frac{C_q h}{g_s e^2} = \frac{\lambda \epsilon_0}{g_s\alpha} \sqrt{\frac{\mu_0}{\epsilon_0}}
\end{equation}

and

\begin{equation}
 \label{eq:tauL}
	\tau_L = \frac{L_q}{R_q} = \frac{\mu_0 \lambda g_s e^2}{4\pi h} = \frac{\alpha \mu_0 \lambda g_s}{2\pi} \sqrt{\frac{\epsilon_0}{\mu_0}},
\end{equation}

\noindent where, as noted, they are functions of the radius $r$ of the orbit. Reference values can be taken by using the Bohr radius $r_b \sim 5.29 \times 10^{-2}$ nm, for which $\tau_C$ and $\tau_L$ are $\sim 7.61 \times 10^{-17}$ and $\sim 2.57 \times 10^{-21}$ s, respectively. These values are the lowest reference electric and magnetic characteristic time constants for an electron ground-state oscillatory orbital motion.

In the following, it will be demonstrated that, by combining Eqs.~\ref{eq:tauC} and \ref{eq:tauL}, it is possible to define $\alpha$ in terms of the electric $\tau_C$ and magnetic $\tau_L$ characteristic time constants, which is a direct consequence of using electrodynamics that complies Maxwell's equations. 

\subsection{\label{sec:alpha} Meaning of the Fine-Structure Constant within Quantum Electromagnetic Rate Theory}

The ratio between $\tau_C$ and $\tau_L$ can now be established by combining Eqs. \ref{eq:tauC} and \ref{eq:tauL}, leading to $\tau_C / \tau_L = 2 \pi / g_s^2 \alpha^2$ and hence providing a meaning for $\alpha$, as a function of $\tau_L$ and $\tau_C$, such as $\alpha = \sqrt{\left( 2\pi / g_s^2 \right) \left( \tau_L / \tau_C \right)}$ or (by noting that $g_s$ equals to 2)

\begin{equation}
 \label{eq:alpha-tau}
	\alpha = \sqrt{\frac{\pi}{2} \left( \frac{\tau_L}{\tau_C} \right)}.
\end{equation}

\noindent This can be alternatively rewritten, as a function of $R_q$, $C_q$, and $L_q$ quantum electromagnetic circuit elements, as

\begin{equation}
 \label{eq:alpha-cir-elements}
	\alpha = \frac{1}{R_q} \sqrt{\frac{\pi}{2} \left( \frac{L_q}{C_q} \right)}.
\end{equation}

\noindent However, perhaps the most insightful meaning of $\alpha$, within this quantum electromagnetic rate theory, is

\begin{equation}
 \label{eq:alpha-sqr}
	\alpha^2 = \frac{1}{g_s^2} \omega_C \tau_L,
\end{equation}

\noindent where $\omega_C = 2\pi /\tau_C = g_s e^2 /\hbar C_q$.

We recall that it is the compliance of the above insights to Maxwell's equations within $1/R_qC_q$ and $R_q/L_q$ rates to describe the electrodynamics of the ground state in an insightful way that enables us to understand the above-defined meaning of $\alpha$. Accordingly, Maxwell's equations will be analyzed in the following subsection in light of the meaning of $\alpha$ and subsequently, the impedance of this closed and electromagnetically induced motion will be analyzed.

\subsection{\label{sec:Maxwelling-equations} Compliance of Quantum Rate Dynamics with Maxwell's Equations}

As demonstrated elsewhere~\citep{Bueno-2023}, Maxwell's equations of classical electrodynamics can be rewritten in a quantum formulation by introducing the above-defined rates $1/R_qC_q$ (electric) and $R_q/L_q$ (magnetic), hence forming the foundation for a quantum electromagnetic rate theory. These laws can be, respectively, stated as

\begin{equation}
 \label{eq:Faraday-law}
	\oint_{\scriptstyle\partial\,\Sigma} \textbf{E} \,d\textbf{l} = - 2 \pi \frac{\Phi_B}{\tau_C} = |\textbf{E}| \lambda
\end{equation}

\noindent and

\begin{equation}
 \label{eq:Maxwell-law}
	\oint_{\scriptstyle\partial\,\Sigma} \textbf{B} \,d\textbf{l} = \mu_0 \varepsilon_0 \frac{\Phi_E}{\tau_L} = |\textbf{B}| \lambda,
\end{equation}

\noindent where $\Phi_B = h/e = \iint_{\partial\,\Sigma}\textbf{B}\;d\textbf{S}$ in Eq.~\ref{eq:Faraday-law} is the magnetic flux quantum and $\Phi_E = \oiint_{\partial\,\mho}\textbf{E}\;d\textbf{S}$ in Eq.~\ref{eq:Maxwell-law} is the electric flux. $|\textbf{E}|$ and $|\textbf{B}|$ are the magnitude of electric and magnetic field vectors $\textbf{E}$ and $\textbf{B}$, respectively. In Eqs. \ref{eq:Faraday-law} and \ref{eq:Maxwell-law}, $|\textbf{E}| \lambda$ and $|\textbf{B}| \lambda$ are the result of the integrand operation over any orbital surface $\Sigma$ with a closed boundary curve $\delta \Sigma$. Note that $\epsilon_i$ = $\oint_{\scriptstyle\partial\,\Sigma} \textbf{E} \,d\textbf{l}$ is the induced electromotive force from the closed induced work $e\epsilon_i$ = $\oint_{\scriptstyle\partial\,\Sigma} \textbf{F}_i \,d\textbf{l}$ (or, equivalently, an induced ground-state energy level component of the electrodynamics) directly accomplished by applying a quantum rate approach for describing the electrodynamics of the ground state~\citep{Bueno-2023}.

Equations \ref{eq:Faraday-law} and \ref{eq:Maxwell-law} can be greatly simplified by noting that $\Phi_E =  -e / \epsilon_0$ (in agreement with Gauss's law) and $\omega_C = g_s e^2 / \hbar C_q$, leading, respectively, to $|\textbf{E}| \lambda = - \omega_C \Phi_B$ and $|\textbf{B}| \lambda = \mu_0 i_L$, where $i_L = - e / \tau_L$ is defined as the electric current associated with the magnetic periodic dynamics expressed in terms of its characteristic time as $\tau_L$, which is induced by electric orbital dynamics with a rate of $\omega_C = g_s g_e \omega_e$ that defines a ground-state energy of $E_{gs} = \hbar \omega_e = e^2 / g_e C_q \sim 13.6$ eV~\citep{Bueno-2023} by taking the Bohr atomic model and the Bohr radius as of referential position in space for the ground-state electrodynamics.

In a previous work~\citep{Bueno-2023}, Eq.~\ref{eq:Maxwell-law} was resolved, straightforwardly leading to Eq.~\ref{eq:h}~\footnote{This was done by considering $\Phi_E$ as $\Phi_E = 2 \lambda \oint_{\scriptstyle\partial\,\Sigma} \textbf{F}_i \,d\textbf{l}$ in Eq.~\ref{eq:Maxwell-law}}. In the following, it will be demonstrated that the combination of Eqs. \ref{eq:Faraday-law} and \ref{eq:Maxwell-law} produces, alternatively, Eq.~\ref{eq:alpha-sqr}. The goal of doing this is to exemplify the suitable implementation of the quantum rate concepts to classical electrodynamics, not only leading to a self-consistent quantum rate analysis but also proving the existence of a quantum phase coherence (associated with the meaning of $\alpha$) between electric and magnetic dynamics. Consequently, the combination of Eqs. \ref{eq:Faraday-law} and \ref{eq:Maxwell-law} leads to

\begin{equation}
 \label{eq:combination-EM-laws}
	c_* = \frac{|\textbf{E}|}{|\textbf{B}|} = \frac{\Phi_B}{\Phi_E} \omega_C \tau_L \left( \frac{1}{\mu_0 \epsilon_0} \right) = \frac{\Phi_B}{\Phi_E} \omega_C \tau_L c^2,
\end{equation}

\noindent where $c^2 = 1 / \left( \mu_0 \epsilon_0 \right)$ is the square of the speed of light, $c$, and $c_*$ is the Fermi velocity of the electron in its orbital ground-state oscillatory motion. Now, by observing that $\Phi_E / \Phi_B = 4 c_*$~\footnote{This results from $\Phi_E = \oiint_{\partial\,\mho}\textbf{E}\;d\textbf{S}$ operating over a closed boundary surface $\partial \mho$ of a volume $\mho$, resulting in an area of $4 \pi r^2$ for the electric flux $\Phi_E$, whereas $\Phi_B = \iint_{\partial\,\Sigma}\textbf{E}\;d\textbf{S}$ is operating over a closed boundary curve of a surface $\Sigma$, resulting in an area of $\pi r^2$ for magnetic flux $\Phi_B$} and $\alpha = c_*/c$, it directly arises that $\alpha^2 = \left( 1/4 \right) \omega_C \tau_L$, \textit{quod erat demonstrandum}. 

Accordingly, the above analysis is not only self-consistent (because $\alpha$ is an essential part of the compliance of quantum rate theory to Maxwell's equations) but  also demonstrates that there is a quantum coherence between electric and magnetic rate dynamics governed by $\alpha$, where $\omega_C \tau_L = g_s^2 \alpha^2$ is a key component of the dynamics, as will be corroborated by an impedance analysis of this oscillatory motion.

In summary, not only do the definitions of $1/R_qC_q$ and $R_q/L_q$ quantum rates suitably comply with Maxwell's equations, but self-consistently these rates inherently provide the meaning of $\alpha$ (or its squared version $\alpha^2$). In the following subsection, the electric impedance of this quantum electromagnetic rate approach will be analyzed.

\subsection{\label{sec:circuit} Quantum Electromagnetic Circuit Analysis of the Ground State}

It is well known that the series combination of inductive $L$ and capacitive $C$ elements in a classical electric circuit analysis leads to an inductive--capacitive ($LC$) nature with an angular characteristic frequency of $\omega_{LC} = 1 / \sqrt{LC}$. In this classical analysis, $\omega$ is always damped by intrinsic contacts and resistances of the circuit and cannot indefinitely sustain its dynamic internal energy oscillation, unless a theoretical idealization is set for perfect $L C$ circuit dynamics that is established in the hypothetical absence of $R$. Therefore, in a classical analysis, the intrinsic presence of resistances as dissipative sources of energy produces a continuous energy loss per cycle of electric oscillating motion. In other words, periodic oscillations in classical $RLC$ circuits are intrinsically nonadiabatic in a way that for the continuous operation of the circuit in an oscillatory external source of energy (preferentially in resonance with the characteristic frequency $\omega$) is required to maintain the electric oscillatory motion. 

An electric impedance $Z$ analysis of a classical series $RLC$ combination of the $R$, $L$, and $C$ elements of the circuit, using a phasorial analysis, provides the square of the impedance as $Z_{RLC}^2 = R^2 + \left( X_L - 1 / X_C \right)^2 = R^2 + \left[ \omega L - 1 / \left( \omega C \right) \right]^2$, where $X_L = \omega L$ and $X_C = 1 / \omega C$ are the inductive and capacitive reactances of the circuit, respectively. The minimum impedance of the circuit corresponds to a maximum electric current $i = V / Z_{RLC}$ flowing internally, where $V$ is the voltage of the circuit. Because the impedance is a function of an external oscillating frequency $\omega$, the minimum impedance occurs for an $\omega$ value that equates to the internal natural characteristic frequency of $\omega_{LC} = 1 / \sqrt{LC}$, which is the case whenever $\omega L = 1 / \omega C$. The system is thus said to be in resonance with the external frequency $\omega$.

Note that in this classical impedance analysis, $Z_{RLC}$ cannot be null because of the inherent existence of dissipative resistive elements, which are constant sources of internal energy loss per cycle of this periodic dynamics. This leads to a constant internal energy loss whenever there is the presence of a classical resistive element. In the ideal hypothetical situation of $R$ being zero, the imposition of a resonant $\omega_{LC}$ would not be physically viable (according to the first law of thermodynamics), because this would lead to an infinite value of $i$. However, in the presence of any source of resistance $R$, the internal energy cannot (without a continuous external feeding of energy) be indefinitely sustained.

In a quantum $RLC$ circuit version of the above-mentioned impedance analysis, the reactances will be defined in agreement with the Faraday and Maxwell inductive laws provided, respectively, by Eqs. \ref{eq:Faraday-law} and \ref{eq:Maxwell-law}. Recall that these equations settled a coherent phase dynamics that led to the meaning of $\alpha$, as provided by Eq.~\ref{eq:alpha-sqr}. According to this description of the dynamics, the inductances can be written as $X_{L_q} = L_q / \tau_L$ and $X_{C_q} = 1 / \omega_C C_q$~\footnote{Note that compared to the classical analysis the angular representation was disregarded in $X_{L_q}$ to comply with $2\pi$ differences of the electric and magnetic rate dynamics as settled in the quantum electromagnetic analysis of equations \ref{eq:Faraday-law} and \ref{eq:Maxwell-law}.}. With the definitions of magnetic $X_{L_q}$ and capacitive $X_{C_q}$ inductances, an impedance analysis can now be similarly conducted on the above-discussed classical situation. The difference here is that classical inductances are replaced by quantum reactance versions, leading to an equivalent analysis of the resonance of the circuit as $\omega_L L_q = 2\pi / \omega_C C_q$, where $\omega_L = 2\pi/\tau_L$. This does not require an external source of energy to sustain its dynamics because of the adiabatic nature of $R_q$, as settled in Eq.~\ref{eq:R_q-magnetic}. The latter statement can be demonstrated using definitions set out in Eqs. \ref{eq:tauC} and \ref{eq:tauL}, where it can be noted that $\tau_C \tau_L = L_q C_q$ in a way that $X_{L_q} - 1 / X_{C_q}$ differences are zeroed in the analysis of $Z_{R_qL_qC_q}^2 = R_q^2 + \left( X_{L_q} - 1 / X_{C_q} \right)^2$, leading to $Z_{R_qL_qC_q} = R_q$.

An analysis of Eqs. \ref{eq:Faraday-law} and \ref{eq:Maxwell-law} and their combination that resulted in Eq.~\ref{eq:alpha-sqr} allows us to infer that it is the electric flux  $\Phi_E$ that sustain the magnetic flux quantum $\Phi_B$ and vice versa in a self-induced orbital dynamics that has its origin in the meaning of $\epsilon_i = \omega_C \Phi_B = \left( 4\alpha^2 /\tau_L \right) \Phi_B = 4 \alpha^2 |\textbf{B}| \lambda c^2$. For instance, noting the meanings of $\alpha = c_*/c$, $\varepsilon_i$, and $|\textbf{B}| \lambda$ in Eqs. \ref{eq:Faraday-law} and \ref{eq:Maxwell-law} leads to $\oint_{\scriptstyle\partial\,\Sigma} \textbf{F}_i \,d\textbf{l} = 4 ec_*^2 \oint_{\scriptstyle\partial\,\Sigma} \textbf{B} \,d\textbf{l}$, where, finally, noting that $F_B = e c_* |B|$ is the magnetic force and $4c_* = \Phi_E / \Phi_B$ (as defined in Section~\ref{sec:Maxwelling-equations}), we obtain

\begin{equation}
 \label{eq:self-inductive-QED}
	\Phi_B \oint_{\scriptstyle\partial\,\Sigma} \textbf{F}_i \,d\textbf{l} = \Phi_E \oint_{\scriptstyle\partial\,\Sigma} \textbf{F}_B \,d\textbf{l}.
\end{equation}

\noindent This expression makes it explicit that there is an electrically induced work $\oint_{\scriptstyle\partial\,\Sigma} \textbf{F}_B \,d\textbf{l}$ (because of the periodic time variations of $\Phi_E$) that creates magnetically induced work $\oint_{\scriptstyle\partial\,\Sigma} \textbf{F}_i \,d\textbf{l}$ (because of a periodic time variations of $\Phi_B$) and vice versa. This electromagnetic dynamics resembles that of the electric and magnetic mutually self-induced wave nature of radiation as described by Maxwell's equations~\citep{ED-Griffiths-1987}, but with an integer difference of $\pi$ between the electric and magnetic phasorial inductive wave phenomena. 

Therefore, $\omega_C \tau_L$ governs this self-induced electrodynamics, where the $2 \pi$ integer difference (for a complete periodic cycle of the motion) is not only a matter of a temporal reference between the electric and magnetic phase dynamics but in fact is important for the analysis of the phenomenon as quantum of space-time.

The consistency of the findings for $Z_{R_qL_qC_q} = R_q$ can now be verified by analyzing the electric current $i = V_q /Z_{R_qL_qC_q}$ provided by the elementary charge of the electron in its angular oscillatory motion associated with $C_q$; that is, $e/\tau_C$. To interpret $i = V_q /Z_{R_qL_qC_q}$, the meaning of the quantum voltage $V_q$ is expected to be consistent with the meaning of $\epsilon_i$. Therefore, assuming that $V_q = \epsilon_i$, we get $V_q = e/C_q$, where $i = e / \left[ \left( Z_{R_qL_qC_q} \right) C_q \right]$ and $Z_{R_qL_qC_q} = R_q$, hence $i = e / R_q C_q = e / \tau_C = i_C$, leading to a self-consistent analysis, \textit{quod erat demonstrandum}.

Finally, the meaning of $R_q$ in the above quantum impedance circuit analysis compared to the classical $R$ needs to be addressed. According to $i_C = \epsilon_i / Z_{R_qL_qC_q} = e / R_q C_q = \epsilon_i / R_q$, the self-induced electric current is quantized by the electromagnetic character of $R_q$, as shown by Eq.~\ref{eq:R_q-magnetic}, and cannot be infinity, as $Z_{R_qL_qC_q} = R_q$ cannot be zero (i.e., it has a quantum limit settled by $Z_0$ and $\alpha$). Consequently, the maximum value allowed for $i_C$ is $\sim$ 2.10 mA~\footnote{This is an astonishing value of electric current given that it is caused by a single electron.}, because of the minimum quantized value of $\tau_C = R_q C_q = \left(h / g_s e^2 \right) 2 \lambda \epsilon_0$ of $\sim 7.61 \times 10^{-17}$ s obtained by taking the Bohr radius as the reference value for the position of the electron in free space.

\section{\label{sec:relativity} Does God Play Dice?}

In this section, the way in which the quantum rate theory correlates with the mass--energy relationship, as originally proposed by Einstein, will be discussed. Let us start by quoting one of Einstein's key statements (from one of his \textit{annus mirabilis} papers~\citep{Einstein-Collection-1987}):

\begin{displayquote}
``If a body gives off energy $E$ in the form of radiation, its mass diminishes as $\left( E/c^2 \right)$."
\end{displayquote}

\noindent In a quantitative physical way, this was established as~\citep{Einstein-Collection-1987}

\begin{equation}
 \label{eq:m-E-relationship}
	\Delta K = \frac{1}{2} \left( \frac{E}{c^2} \right) v^2,
\end{equation}

\noindent where $\Delta K$ is the variation of kinetic energy upon radiative emission of an amount of energy $E$, $v$ is the velocity of the body, and $E/c^2$ is the lost mass during the irradiation of energy by the body, as clearly pointed out in the above-quoted statement.

Following symmetric physical reasoning to that of Einstein's statement raises the following question:

\begin{displayquote}
``If an electron (as a specific type of body with an electromagnetic character) in its ground-state energy level absorbs radiation energy of $E = hc / \lambda$, would it not increase its mass by $\left( E / c^2 \right)$?"
\end{displayquote}

The goal of the present section is to analyze this question in light of a quantum electromagnetic rate interpretation. Let us start by noting that $\alpha$ can be described as a function of $C_q$. This can be made by combining and rearranging Eqs.~\ref{eq:h} and \ref{eq:C_q-GS}, thus leading to $\alpha = \left( e^2/C_q \right) \left( \lambda / hc \right)$, where $e^2 / C_q = g_e E_{gs} = e \epsilon_i = h \nu_e$ and $\lambda / hc = 1 / E$. This analysis directly enables us to establish a relationship between $E$ and $E_{gs}$ such as $\alpha E = g_e E_{gs}$.

Now, if we multiply both sides of Eq.~\ref{eq:alpha-sqr} by half of the energy of the absorbed photon, $\left( 1/2 \right) E = \left( g_e/ 2\alpha \right) E_{gs}$, and rearrange the equation, this leads to
\begin{equation}
 \label{eq:K-pe}
	 \Delta K_{pe} = \frac{1}{2} \left( \frac{E}{c^2} \right) c_* = \frac{g_e E_{gs}}{8 \alpha} \omega_C \tau_L = \frac{g_e E_{gs}}{2},
\end{equation}
\noindent where $\Delta K_{pe}$ is identified as the variation of the kinetic energy of the electron (or of the photon--electron entity created) upon the absorption of the photon by the electron. Observing that $g_e = 2$ leads to $\Delta K_{pe} = E_{gs}$, in agreement with the traditional quantum mechanical interpretation of a photon absorption event by an electron in its ground-state level of energy. 

However, in terms of a quantum electromagnetic rate interpretation of this event, $E_{gs}$ equates to $e \epsilon_i = e^2 / g_e \hbar C_q = \hbar \omega_e$, providing additional insight. For instance, although $\Delta K_{pe} = E_{gs}$ is the expected outcome from traditional quantum mechanics (capable of explaining the experimental results), it is only by means of alternatively denoting this variation of energy as $\left( g_e E_{gs} / 8 \alpha \right) \omega_C \tau_L$ that the role played by $\alpha$ is taken into account. Through the mathematics inherently existing in traditional quantum mechanics interpretation, the meaningful physics is lost (with uncertainties), but the interplay between $\alpha$ and $h$ comes for the interpretation of a photon absorption event by an electron in the ground state. It is only by considering an electromagnetic rate character of the event that the essential roles played by $\alpha$ and $h$ can be appropriately taken into account, as discussed in the following subsection.

\subsection{\label{sec:relativistic-momentum} Role of $\alpha$ and Dirac Relativistic Momentum}

Additionally, the role of $\alpha$ in the ground state can be analyzed by establishing a relationship between the mass $m_c$ of the electron and the magnetic components of the motion. This relationship can be taken by considering the magnetic component of the Lorentzian centripetal force. In doing this, it can be observed that $\omega_C$ equates to $e |\textbf{B}|/m_c$, where $m_c$ is the synchrotron mass of the electron~\cite{Novoselov-2005}. This allows us to analyze the particle--wave duality with $\alpha$ as a parameter in the equations. Trivial algebraic operations lead to three {meaningful} relationships:
\begin{equation}
 \label{eq:E-tauL}
	 E \tau_L = \frac{e^2}{2 \pi} \sqrt{\frac{\mu_0}{\epsilon_0}} = 2 \alpha \hbar,
\end{equation}
\begin{equation}
 \label{eq:alpha-h-meaning}
	 \alpha h = \frac{e^2}{g_e} \sqrt{\frac{\mu_0}{\epsilon_0}} = \frac{e^2}{g_e} Z_0,
\end{equation}
\noindent and the linear momentum $p = m_c c_*$ expressed as
\begin{equation}
 \label{eq:momentum}
	 p = \frac{\mu_0}{2 L_q} \hbar,
\end{equation}
\noindent where all these three relationships were obtained by trivial algebraic manipulations of $\omega_C = e |\textbf{B}|/m_c$.

Equation~\ref{eq:E-tauL} demonstrates that the mathematical product of $\alpha$ and $\hbar$, within an electromagnetic meaning based on the reinterpretation of $h$ provided by Eq.~\ref{eq:h}, equates to the product of $E$ of the photon and $\tau_L$ of the electron. Therefore, ignoring the interaction of the photon with the magnetic dynamics accomplished by $\tau_L$ (in the orbital dynamics of the electron) introduces an uncertainty provided by the mathematical product of $\alpha$ and $h$ (with physical units of joule-seconds). 

The reintroduction of the meaning of $E = g_e E_gs / \alpha$ in Eq.~\ref{eq:E-tauL}, which leads to $g_e E_{gs} \tau_C = h / 2$, is consistent with the phenomena observed from an electric field hypothetical ``experimental'' perspective. Note that, because the uncertainty principle~\citep{Sen-2014} requires $\Delta E \Delta \tau \geq h/2$ for this ``hypothetical'' experiment, the obtained uncertainty is incremented by $g_e$ over the ``measurement''. This increases by an exact ``uncertainty'' of $g_e$ for each experiment owing to the loss of information introduced as a result of disregarding the induced electric component during this ``experiment'' on the dynamics. 

In other words, the interaction of the energy $E$ of a hypothetical photon absorbed by an electron at its ground-state energy of $E_{gs}$, as measured from an electric field perspective of the photon, introduces ``uncertainties'' that can be now understood in terms of a byte of information lost during the observation of the ``experiment.'' This is a consequence of the interplay between electromagnetic and mechanical energy during the instant of the absorption of the photon by the electron, with the photon only interacting with the electromagnetic component as will be discussed below.

Equation~\ref{eq:alpha-h-meaning} reinforces the analysis just described by confirming that the mathematical product between $\alpha$ and $h$ is purely electromagnetic, in agreement with the interpretation of Eq.~\ref{eq:h}. Additionally, this equation demonstrates that ignoring the inherent electric degeneracy of the electron have consequences for the interpretation of quantum electrodynamics, which can be envisaged by comparing Eq.~\ref{eq:alpha-h-meaning} with Eq.~\ref{eq:h}. Essentially, both equations are numerically equivalent, but, in the latter, $g_e$ is explicitly demonstrated in the meaning of the product of $\alpha$ and $h$, indicating a ``change in the phase of the wave during the reflection'' of the photon by the electron in this ``experiment".

Finally, Eq.~\ref{eq:momentum} is the linear momentum (and thus the left side of this equation describes a particle behavior) of the electron in the orbit within a wave electromagnetic massless representation (as expressed on the right side). The result, as expected, is consistent with the traditional quantum mechanics interpretation of wave--particle duality, which, by substituting the definition of the ground-state inductance quantum $L_q$, as stated in Eq.~\ref{eq:L_q-GS}, into Eq.~\ref{eq:momentum} leads to $\mu_0 / 2 L_q = 1 / r = 2\pi / \lambda = |\textbf{k}|$. This demonstrates that Eq.~\ref{eq:momentum} not only complies with the de Broglie $p = h \lambda$ wave--matter interpretation of this electrodynamics phenomenon (as an example of wave--particle duality), but it also settled the electromagnetic meaning for the Dirac wavevector $\textbf{k}$~\citep{Dirac-1928}. Consequently, Eq.~\ref{eq:momentum} is consistent with the Dirac equation~\citep{Dirac-1928} in which the massless Fermionic nature of the electron, as observed in graphene through a quantum Hall experimental analysis~\cite{Novoselov-2005}, is an intrinsic characteristic of the theory. The quantum electromagnetic rate theory as stated here demonstrates that the massless nature of the electron results from the inductive magnetic nature of the dynamics. In graphene, this massless nature of the electrodynamics is observed only at the Dirac point~\cite{Novoselov-2005}, where the quantum rate was noted to be maximum~\citep{Bueno-QRGraphene-2022}. Therefore, quantum electromagnetic rate theory not only complies with these results but it provides additional physical information to understand the electrodynamics of graphene. 

In summary, the role played by $\alpha$ has not been appropriately considered within the meaning of $h$ or of quantum mechanics. This is a result of the phase coherence between electric and magnetic temporal dynamics that has not been properly considered so far.

\subsection{\label{sec:relativistic-momentum} Is There Really an Uncertainty?}

All the analyses conducted so far, within a quantum electromagnetic rate interpretation of the ground-state energy~\citep{Bueno-2023}, come from the initial hypothesis of the existence of an additional and magnetically induced  electric potential energy $e\varepsilon$. Note that $e\varepsilon_i$ is, in magnitude, comparable to that considered by Bohr, that is, with $E_{gs} = e\varepsilon_i = e^2/g_e h C_q \sim 13.6$ eV. Nonetheless, the electric potential energy considered in the Bohr model has a different origin associated with Gauss's law, in which the electric field comes from the elementary charge $e$ of the electron \textit{per se} and not from its dynamics as predicted by Faraday's induction law that leads to $\varepsilon_i$. 

The Planck constant $h$ was introduced in Bohr theory in an \textit{ad hoc} way and was associated with the ``electrostatic'' potential energy $E_e$ interpretation of the mechanics of the phenomenon. The analysis led by Bohr and further by Schr\"{o}dinger (both in agreement with de Broglie’s wave--matter hypothesis) conducted to develop nonrelativistic quantum mechanics cannot handle the magnetic component of the electron (requiring further consideration of this component as an energy degeneracy such as $g_s$). Accordingly, a nonrelativistic quantum mechanics description of the electrodynamics requires consideration of the kinetic energy $K = m_0 c_*/2$ computed as half of $E_e$, requiring the implementation of a $g_e$ degeneracy. 

Later, the development by Dirac~\citep{Dirac-1928} and Feynman~\citep{Feynman-1949} of quantum electrodynamics formulated the problem from another (relativistic) perspective  in which the magnetic character of the electron within its spin component is considered differently. In Dirac's approach, the spin was introduced as a mathematical entity~\citep{Dirac-1928}; in Feynman's approach~\citep{Feynman-1949, Feynman-1985}, $\alpha$ is required to sustain the theory, with a different electromagnetic perspective, but none of these theories predicted an electric degeneracy $g_e$ of the ground state. Due to ignoring $g_e$, these theories could not reconcile quantum mechanics with Maxwell's electromagnetic theory from which Einstein formulated the theory of relativity. 

Dirac and Feynman formulations of quantum electrodynamics lack physical meaning by requiring knowledge of the role played by $\alpha$~\cite{Feynman-1985} or by the need to interpret what the mathematical description of a physical problem is telling us about its nature~\citep{Dirac-1928}. It is remarkable that the ground-state quantum mechanics was developed so far by ignoring an equivalent energy component $e\epsilon_i$ of the dynamics that is equivalent in magnitude to that of $E_e$ but that has a  different physical nature associated with the inductive dynamics predicted by Faraday's law. This $e\epsilon_i$ source of electric potential energy can coherently interact with a photon or be perturbed by other types of electromagnetic stimuli, such as an oscillatory electric field in electrochemical experiments~\citep{Bueno-QR-foundation-2020, Alarcon-2022} or a magnetic field in a quantum Hall analysis of electrodynamics in graphene~\cite{Novoselov-2005, Bueno-QRGraphene-2022}. 

Introducing this magnetically induced energy $e\varepsilon_i$ into quantum mechanics enables a reinterpretation of the electrodynamics at the ground-state energy level $E_{gs}$. It can now be noted that an \textit{ad hoc} consideration of $g_s$ is unnecessary in a quantum electromagnetic rate interpretation of the ground-state energy phenomenon. This is because there is a fundamental electric field degeneracy $g_e$ that has its origin solely in $e$ that leads to two distinct electric field contributions to the electrodynamics that complies with Maxwellian dynamics, which includes the magnetic dynamics. 

Accordingly, it is possible to reinterpret the origin of the magnitude of $E_{gs}$ degeneracy as $E_{gs} = E_e + e\varepsilon_i \sim 27.2$ eV, where it is assumed that the magnitude of the total electric potential energy can be stated equivalently as $g_e E_e \sim 27.2$ eV or $g_e e\varepsilon_i \sim 27.2$ eV and where $g_e = 2$ conforms with an energy degeneracy in which the particle $E_e = e^2/g_e4\pi \varepsilon_0 r \sim 13.6$ eV is equivalent to the wave $e\varepsilon_i = e^2/2 g_e \lambda \varepsilon_0 \sim 13.6 $ eV~\citep{Bueno-2023}) nature. This promotes a degeneracy with two distinct physical origins, in which the wave and particle states coexist. The above analysis provides an insightful interpretation of the interaction of the ground-state electrodynamics with a photon of energy $E$. 

The following analysis demonstrates that the particle or wave nature of the electron is revealed depending on which of these coexisting electric field states are observed, depending on a perspective of $E_e$ (owing to Gauss's law) or $e\varepsilon_i$ (owing to Faraday's law), both with the same magnitude, but with different electrodynamics laws governing their meanings, permitting a coexisting wave and particle nature of the electron, in which a magnetic coherent dynamics governed by $\alpha$ entraps a charged particle with a $m_0$ resting mass.

The observed reference values of these states promotes a particle or a wave interpretation of the phenomenon that is not involved with uncertainties, but only with the relativity of the event. This simple reinterpretation of the origin of the energy degeneracy of the electron not only complies with previous quantum mechanics analysis, as demonstrated here, but has profound physical consequences for the interpretation of particle--wave duality. 

For instance, it implies that the electric charge $e$ and the resting mass component $m_0$ of the electron particle--wave dynamics acquires a relativistic character with an energy of $E_e =  m_0 c_*^2$. This energy contributes additionally to that of $e\varepsilon_i$, with $E_e$ and $e\varepsilon_i$ depending only on the ``static'' (resting component associated with $E_e$) or dynamic (associated with $e\varepsilon_i$) energy sources of a force field $e|\textbf{E}|$, respectively, with a relativistic energy computed as $E_{gs} = \sqrt{\left( e\varepsilon_i \right)^2 + \left( m_0 c_*^2 \right)^2}$. This can be interpreted as if the electromagnetic dynamics of a charged particle, at a particular space--time setting (referred to as the ground state), can entrap its resting mass $m_0$ in a self-induced electromagnetic dynamics with an inherent relativistic character in which $\alpha$, as the ratio between the electric and magnetic rates, plays a prominent role for the sustainability of this particle--wave dynamics.

Assuming now that the electron, in this electric field degeneracy state, with an energy of $\sim$27.2 eV, can coherently only absorb a photon by means of its magnetically induced electric field component, with a corresponding induced electromagnetic energy of $e\varepsilon_i$, hence the absorption of a photon occurs associated with the absorption of an amount of $e\varepsilon_i/2 \sim 13.6$ eV, which is numerically equivalent to the half of $E_{gs} \sim$ 27.2 eV and that can be accounted as if it was a particle or wave, depending on the reference observations, that is that of the absorbed energy (wave) or of the emitted (particle) during a photon-absorption perturbation of the electromagnetic coherence of ground-state $e\epsilon_i$ component of $E_{gs} \sim$ 27.2 eV. 

Consequently, this interplay can be considered as a ``classical'' mechanical (or particle) physical event of a quantum wave mechanical absorption phenomenon, from which the uncertainty originates as a loss of half of a byte of information. Note also that this small amount of emitted energy can be interpreted as having a kinetic character because $\Delta K_{pe} = m_0 c_*/2 \sim 13.6$ eV, as predicted by Eq.~\ref{eq:K-pe}, if observed from the emission perspective.

As a result, the absorption of the photon energy of $\sim$13.6 eV by the electron at a ground-state energy $e\varepsilon_i = \oint_{\scriptstyle\partial\,\Sigma} \textbf{F}_i \,d\textbf{l}$ requires consumption of energy for releasing the electron from its inherent electromagnetic phase coherence governed by the work $\oint_{\scriptstyle\partial\,\Sigma} \textbf{F}_i \,d\textbf{l}$, with the remaining energy from half of $m_0 c_*$. Therefore, the consumption of $e\varepsilon_i = h\nu_e \sim 13.6$ eV is required to release the electron from its electromagnetic phase coherence $\oint_{\scriptstyle\partial\,\Sigma} \textbf{F}_i \,d\textbf{l}$ with the emission of energy $\Delta K_{pe} = m_0 c_*/2 \sim 13.6$ eV, which is the classical kinetic energy of a free electron. 

In other words, upon the absorption of an amount of energy $e\varepsilon_i/2 \sim 13.6$ eV in the form of electromagnetic radiation, an electron with charge $e$ and resting mass $m_0$ is released (emitted) from its entrapped coherent electromagnetic dynamics within the atom, as governed by compliance with an electromagnetic constant $\alpha$, to freely travel with half of its energy of $e\varepsilon_i/2 \sim 13.6$ eV, but following a particle behavior because of the remaining energy $E_e = m_0 c_*^2 / 2$, which is an amount equivalent to $\Delta K_{pe}$ (in connection with $\alpha$ as shown in Eq.~\ref{eq:K-pe}), in a dynamics that obeys the energy conservation principle, as required. 

Furthermore, this interpretation of the quantum electrodynamics (in which the wave-particle duality is interpreted as an electro-mechanical coherent phenomenon) is consistent with modern experiments on conductivity and electron transport, as observed at the nanoscale~\cite{Imry-Landauer-1999}. For instance, it can be observed that a classical resistance $R$, as a drifting transport of free electrons by means of an externally driven difference of potential that operates over a charge $e$ in metallic structures, has an intrinsic dissipative character. Consequently, it is nonadiabatic, during the process of transport of the electron, as is required by the dynamics of massive objects. Therefore, this classical way of transporting electrons is electromagnetically incoherent whereas the transport of electrons in quantum channels, where a quantum resistance $R_q$ governs the phenomenon~\cite{Imry-Landauer-1999, Sanchez-QR-Rct}, has a coherent and an almost adiabatic pattern, as demonstrated by the electrodynamics of graphene~\cite{Novoselov-2005, Bueno-QRGraphene-2022} and of electrons transported through electrochemical junctions in electrochemical reactions~\citep{Alarcon-2022, Sanchez-QR-Rct, Sanchez-QR-efficience-2022}.

The above analysis and interpretation of the uncertainty principle are consistent with those conducted in agreement with Eq.~\ref{eq:K-pe}. Moreover, the wave--particle duality of the electron (matter) and photon (wave) are explained, as the uncertainties are introduced only if the induced time variation of the magnetic flux and $\alpha$, as the origin of $e\varepsilon_i$, are disregarded in the analysis of the electrodynamics. Uncertainties are introduced in the analysis by the loss of multiples of a half of a byte of quantum information (accounted for as $g_e \times$ the number of experiments) and the wave or particle interpretation of the experiment depends on a relative perspective of observation of the phenomenon, as discussed in Section~\ref{sec:relativistic-momentum}, where the uncertainty principle was interpreted from a quantum electromagnetic rate perspective. 

\section{\label{sec:conclusion} Conclusions}

Quantum electromagnetic rate theory considers that the origin of the degeneracy of electron energy in its ground state is inherently associated with two ways that classical electrodynamics predicts electric fields to exist and that are able of operating over the electrodynamics. This leads to a plausible way (with uncertainties accounted for as multiple of the degeneracy per observation or perturbation of the electromagnetic coherent state) of interpreting the uncertainty principle and energy conservation within this quantum rate electrodynamic approach.

Considering appropriate electric and magnetic quantum rates for describing the electric and magnetic components of the electron in a closed oscillatory motion electromagnetically coupled to a nucleus makes it possible to describe the meaning of $\alpha$ within this quantum electromagnetic theory. The temporal electromagnetic coherence of the orbital motion conforms with both Dirac relativistic quantum and classical Maxwellian electrodynamics that conducts a reinterpretation of the meaning of the constant $h$. It was inferred that the perturbation of the electromagnetic coherence (that couples electrons to its ground-state dynamics) by a photon release an amount of mechanical energy (as a kinetic energy variation) that is equivalent to that absorbed by the electromagnetically induced field component of the electron, thus revealing the particle behaviour of the electron as a free (not electromagnetically coupled) entity.

\bibliography{references}
\end{document}